\documentclass[%
 reprint,
 superscriptaddress,
 twocolumn, 
 amsmath,amssymb,
 aps,
]{revtex4-2}

\usepackage{todonotes}
\usepackage[colorlinks = true,
            linkcolor = blue,
            urlcolor  = blue,
            citecolor = blue,
            anchorcolor = blue]{hyperref}

\usepackage{graphicx}
\usepackage{dcolumn}
\usepackage{bm}
\usepackage{braket}
\usepackage{glossaries}
\usepackage{float}
\newacronym{CDW}{CDW}{charge-density-wave}
\newacronym{QMC}{QMC}{charge-density-wave}
\newacronym{DFPT}{DFPT}{density functional perturbation theory}
\newacronym{DFT}{DFT}{density functional theory}

\usepackage{ulem}

\newcommand{\COMMENTED}[1]{}

\makeatletter
\newsavebox{\@brx}
\newcommand{\llangle}[1][]{\savebox{\@brx}{\(\m@th{#1\langle}\)}%
  \mathopen{\copy\@brx\kern-0.5\wd\@brx\usebox{\@brx}}}
\newcommand{\rrangle}[1][]{\savebox{\@brx}{\(\m@th{#1\rangle}\)}%
  \mathclose{\copy\@brx\kern-0.5\wd\@brx\usebox{\@brx}}}
\makeatother

\begin{document}

\preprint{}

\title{In search of exotic pairing in the Hubbard model: \\
	many-body computation and quantum gas microscopy}

\author{Chunhan Feng}
\affiliation{Center for Computational Quantum Physics, Flatiron Institute, 162 5th Avenue, New York, New York 10010, USA}

\author{Thomas Hartke}
\affiliation{Department of Physics, MIT-Harvard Center for Ultracold Atoms, and Research Laboratory of Electronics,
Massachusetts Institute of Technology, Cambridge, Massachusetts 02139, USA}

\author{Yuan-Yao He}
\affiliation{Institute of Modern Physics, Northwest University, Xi’an 710127, China}
\affiliation{
Shaanxi Key Laboratory for Theoretical Physics Frontiers, Xi’an 710127, China}
\affiliation{Peng Huanwu Center for Fundamental Theory, Xi'an 710127, China}

\author{Botond Oreg}
\affiliation{Department of Physics, MIT-Harvard Center for Ultracold Atoms, and Research Laboratory of Electronics,
Massachusetts Institute of Technology, Cambridge, Massachusetts 02139, USA}

\author{Carter Turnbaugh}
\affiliation{Department of Physics, MIT-Harvard Center for Ultracold Atoms, and Research Laboratory of Electronics,
Massachusetts Institute of Technology, Cambridge, Massachusetts 02139, USA}

\author{Ningyuan Jia}
\affiliation{Department of Physics, MIT-Harvard Center for Ultracold Atoms, and Research Laboratory of Electronics,
Massachusetts Institute of Technology, Cambridge, Massachusetts 02139, USA}

\author{Martin Zwierlein}
\affiliation{Department of Physics, MIT-Harvard Center for Ultracold Atoms, and Research Laboratory of Electronics,
Massachusetts Institute of Technology, Cambridge, Massachusetts 02139, USA}

\author{Shiwei Zhang}
\affiliation{Center for Computational Quantum Physics, Flatiron Institute, 162 5th Avenue, New York, New York 10010, USA}

\COMMENTED{
\author{A,B,C,D,E}
}

\date{\today}

\begin{abstract}
Pair density waves and exotic superconductivity have long been of strong interest, and have attracted much recent attention. We present a joint theoretical and experimental exploration of possible signatures of fermion pairing with finite center-of-mass momentum, or the Fulde-Ferrell-Larkin-Ovchinnikov (FFLO) order. Experimentally a doped and spin-imbalanced attractive two-dimensional Hubbard model is realized with a cold atomic gas in an optical lattice, and quantum gas microscopy is used to probe its properties. Computationally we study the same model with state-of-the-art constrained-path (CP) auxiliary field quantum Monte Carlo (AFQMC). Direct comparisons between experiment and computation on various short-range magnetic and charge correlations show excellent agreement. We then
investigate these correlations, as well as pairing correlation functions, systematically to low 
temperatures with CP-AFQMC, and determine parameter regimes in density and magnetization where signatures of FFLO order may be observed. We show that the temperature at which precursors of such orders appear is already within reach of the current experiment. We discuss routes for direct experimental detection and measurements of such states.

\end{abstract}

\maketitle

\section{Introduction} 

It is rare to have systems with unconventional
superconducting states
for which 
accurate, {\it ab initio\/} computations can also be performed. 
Rarer still is to find a case where the system is also directly accessible by experiment.
The attractive Hubbard model provides one such possibility.
When there is no spin imbalance, 
this model is amenable to exact computations by quantum Monte Carlo \cite{Bertaina2011, Shi2015, Jensen2020, He2022,Scalettar1989,Yuanyao2025}, and experimentally it has been realized with the use of optical lattices \cite{Bloch2008, Giorgini2008,Mitra2018,Gall2020,Chan2020,Brown2020,Hartke2023}. 
Systematic comparisons between computation and experiment over 
the pairing states and the BEC-BCS crossover in a continuum Fermi gas has shown impressive agreement~\cite{Carlson2011-3DFG-unitarity,Ku2012-FG3D,daix2025observingspatialchargespin}. 
In principle similar comparisons are possible on a lattice, as in the repulsive case~\cite{Xu2025,Chalopin2024}, although there has been limited experimental work in the attractive Hubbard model~\cite{Hartke2023,Mitra2018,Gall2020,Chan2020,Brown2020}. 
When there is spin imbalance, the system is more challenging for both computation and experiment. In this paper, we show that the  
spin-imbalanced model 
provides 
a great opportunity for a synergistic approach between experiment and computation, where 
a variety of interesting questions involving correlations and  
pairing can be explored.

One form of exotic superfluid pairing occurs in 
the 
Fulde-Ferrell (FF) \cite{FF1964} and Larkin-Ovchinnikov (LO) \cite{LO1964} phases.
For example, in an unequal population 
of spin up and down fermions
with attractive interactions,
the mismatch between the Fermi surfaces of the two spin flavors can lead to a condensation of pairs with finite center-of-mass momentum $\vec{Q}$,
with pairing order parameters
$\Delta_{\rm FF}({\vec{r} })=\Delta\,\exp(i\,{\vec{Q}}\cdot {\vec{r}})$ 
or  
$\Delta_{\rm LO}({\vec{r}})=\Delta\,\cos({\vec{Q}}\cdot {\vec{r}})$. 
A possible realization is in correlated materials, where there has been strong interest in pair density waves 
\cite{Hamidian2016,Edkins2019,Aishwarya2023,Liu2023,Gu2023a}.
Cold atoms provide a compelling possibility for a clean realization---with finely tunable system parameters---in which two hyperfine states can be viewed as two spin species.

There has been a great deal of effort
to 
search for the FFLO phases
in the past several decades. 
Experimentally, the realization of superfluidity in atomic Fermi gases~\cite{Ketterle2008, Inguscio2008, Giorgini2008,Zwerger2012} enabled the investigation of superfluidity in the presence of spin-imbalance~\cite{Zwierlein2006,Partridge2006a,Shin2006,Schunck2007a,Shin2008}. The observation of the Pauli limit of superfluidity~\cite{Zwierlein2006}, where the chemical potential difference of the two species overcomes the pairing gap, pointed to the region in parameter space where to search for the FFLO phase. However, in trapped gases in 3D~\cite{Zwierlein2006,Partridge2006a,Shin2006,Schunck2007a,Shin2008} and 2D~\cite{Mitra2016}, phase separation between the balanced superfluid core and a spin-imbalanced Fermi liquid of polarons~\cite{Chevy2006a,Prokofev2008a,Schirotzek2009,Massignan2025} so far precluded the observation of an intervening FFLO phase.
Spin-imbalanced Fermi gases in one dimension have been realized~\cite{Liao2010}, but signatures of FFLO pairing have yet to be found.
Fermi gases in box potentials~\cite{Mukherjee2017,Navon2021box} may offer a new route of investigation in any dimension as the density imbalance is homogeneous, allowing to narrow in on the Pauli critical region.

The search computationally
has included a variety of 
 studies 
on  model Hamiltonians
\cite{Castorina2005,Sheehy2006,Kinnunen2006,Machida2006,Parish2007,Hu2007,He2007,Koponen2006,Liu2008,Kakashvili2009,Koponen2008,Loh2010,Heikkinen2014,Dao2008,Chiesa2013,Rosenberg2015,Casula2008,Batrouni2008,Batrouni2009,Wolak2010,Wolak2011,Wolak2012,Gukelberger2016,Foo2019,Vitali2022,Rizzi2008,Tezuka2008,Potapova2023}.
Except for the cases in one dimension (1D) or quasi-1D,  
most of these calculations 
involve significant approximations or uncertainties. 
Direct and predictive computations are challenging, because 
the state in question is the outcome of a delicate 
balance of strong interaction and fermiology that 
require accurate computations in sufficiently large systems and low temperatures. 
While there has been a growing effort to search and study FFLO order in  electronic materials \cite{Bianchi2003,Kakuyanagi2004,Kumagai2006,Agosta2017,Kasahara2020}, it has remained elusive in a simple, clean system that is experimentally accessible.

The two-dimensional (2D) attractive Hubbard model provides a unique platform for  studying the FFLO order. 
On a square lattice it can be mapped to the repulsive Hubbard model via a partial particle-hole transformation (PHT)~\cite{scalettar2016,Xie2025}. 
The latter has been intensely studied, especially in the context of cuprate superconductivity. Despite 
decades of effort, 
much remains unknown about the phase diagram of this model.
Recent progress indicates that FFLO order exists in the attractive Hubbard model at low 
temperatures, at least for some parameter regime (density, magnetization, and attraction strength).
This includes calculations in 
the attractive model with  both diagrammatic Monte Carlo \cite{Gukelberger2016} and auxiliary-field quantum Monte Carlo (AFQMC) \cite{Vitali2022}. Furthermore, a large number of studies in the doped  repulsive Hubbard model have shown the existence of 
spin-density waves and spin stripes \cite{Chang2010,White1998,Zheng2017,Xu2022}, 
which via particle-hole transformation
would also suggest that FFLO order can be present in the attractive model at half-filling.

Experimentally the Hubbard model can be emulated by 
cold atoms in an optical lattice. Cold atom experiments provide access to clean systems where the interaction between the different fermion species can be
tuned via Feshbach resonances~\cite{Gross2017}. Both repulsive ~\cite{Chalopin2024,Xu2025} and attractive ~\cite{Hartke2023,Mitra2018,Gall2020,Chan2020,Brown2020} Hubbard models have been realized.
In the repulsive Hubbard model, antiferromagnetism has been experimentally observed~\cite{Hart2015,Parsons2016,Cheuk2016}, while non-local pairing has been reported in the attractive regime~\cite{Hartke2023}.
One of the experimental challenges has been to lower the temperature sufficiently so as to reach the targeted phases, although encouraging recent progress has been achieved~\cite{Xu2025,Chalopin2024}. 
With direct and quantitative information
from
computation as guidance, 
major steps forward can be expected for the search and experimental realization of 
FFLO, as we hope to demonstrate in the paper.

Computationally, the methods available to perform accurate calculations in the attractive Hubbard model have also seen significant improvement recently.  
The progress 
is in large 
part driven by a community effort in benchmarking \cite{Qin2016_benchmark}
and in multi-method collaborations \cite{LeBlanc2015} to study the repulsive Hubbard model. 
Many of the methods can be 
generalized to the attractive 
model. Indeed some of them even toggle between the two models in practice \cite{Qin2020,Xu2023}, taking advantage of the PHT. However,
most of the studies in the repulsive Hubbard model have focused 
on the doped, spin balanced case,  
which corresponds to the 
spin-imbalanced attractive Hubbard model
at half-filling only.

Here we leverage the advances in both computation and cold atom experiment to study 
the 2D 
attractive Hubbard model. 
The model is realized 
using a gas of fermionic $^{40}$K
in an optical lattice,
and various short-range correlations are measured
using quantum gas microscopy~\cite{Hartke2023}. In conjunction, the same model is studied 
with the
finite temperature (FT) auxiliary field quantum Monte Carlo
method \cite{Zhang1999,He2019}.
Detailed 
comparisons are made between experiment and computation under similar conditions. 
Quantitative agreement is seen 
as a function of density and magnetization 
at temperatures down to the lowest currently accessible by experiment.
We then perform computations to lower temperatures and
systematically study the 
evolution of 
different 
pairing and other correlation functions and properties. 
We show that FFLO order can be expected in large regions of the parameter space (density and doping). Furthermore, we find parameter regions where  
precursors for  
such order 
actually 
appear  
at rather high temperatures $T/t \sim 1/3$, which are  within reach of the current setup of cold atomic gas experiments. We suggest ways to  
use spin correlation functions as proxies to experimentally detect the evolution of FFLO order.
This work illustrates 
a path forward for realizing exotic pairing and other
quantum phases of matter, in which 
the synergy from coupling computation and experiment can greatly 
accelerate progress and breakthroughs.
 
The remainder of the paper is organized as follows. We discuss the Hamiltonian and give an overview of methodology
and experimental setup
in section \ref{sec:System and Overview of Methodology}. Results are presented in section \ref{sec:Results}. 
 These are divided into 
 three subsections focusing, respectively, on a detailed
 comparison between experiment and computation, temperature dependence and systematic results at lower temperatures from computation, and 
 a final ``phase diagram'' of FFLO together with a proposal for experimental detection.
 We then conclude and give a brief outlook in 
 section \ref{sec:Discussion}.

\section{System and Overview of Methodology} 
\label{sec:System and Overview of Methodology}
We consider
the doped, spin-imbalanced, attractive Hubbard model on a square lattice. The system is defined by the Hamiltonian
\begin{align}\label{eq:Hubbard_N1}
\nonumber
\hat H = & -t \sum_{\langle i,j \rangle, \sigma} \left( \hat c_{i \sigma}^\dagger \hat c_{j \sigma}^{\phantom{\dagger}} + \mathrm{h.c.} \right) + U\sum_{i} \hat n_{i \uparrow} \hat n_{i \downarrow} \\ 
& 
-\mu\sum_{i} \left(\hat n_{i \uparrow}+ \hat n_{i \downarrow}\right)
-\frac{B}{2}\sum_{i} \left(\hat n_{i \uparrow}- \hat n_{i \downarrow}\right)\,,
\end{align} 
where $\hat c_{i \sigma}^\dagger$ ($\hat c_{i \sigma}^{\phantom{\dagger}} $) creates (annihilates)  
a fermion with spin $\sigma$ 
($\uparrow$ or $\downarrow$) on site $i$,
and the density operator is $\hat n_{i\sigma}\equiv \hat c_{i \sigma}^\dagger \hat c_{i \sigma}$.
The first term 
in Eq.~(\ref{eq:Hubbard_N1})
describes the
hopping process between nearest neighbors $\langle i,j \rangle$. The next term denotes the 
attractive on-site interaction, with strength $U<0$. 
The chemical potential $\mu$ and magnetic field
$B$ (along the $z$-direction)
are coupled to the charge and spin operators respectively:
$\hat n_i\equiv 
\hat n_{i \uparrow}+ \hat n_{i \downarrow}$ and
$\hat S^z_i\equiv 
(\hat n_{i \uparrow}- \hat n_{i \downarrow})/2$. 
In this work, we focus on intermediate interaction strengths $U\sim -4t$. With no loss of generality, $B$ is chosen to be positive, hence spin $\uparrow$ fermions are the majority fermions. 
The Hamiltonian can be rescaled by $t$, which defines the the energy 
units and in which we measure all 
the energy scales.

In the experiment the external trapping potential controls the local chemical potential $\mu(\vec{r}) = \mu_0 - V(\vec{r})$, allowing the simultaneous measurement of density, magnetization, and correlations over a range of chemical potentials~\cite{Hartke2020}.
Our computations use either a square or rectangular simulation cell, for different purposes as elaborated below. The simulation cell size will be denoted 
by $N=L_x \times L_y$.
Periodic boundary conditions (PBC) are applied  
in both $x$ and $y$ directions.

We characterize the system by various observables and correlation functions. For example the overall density and magnetization,
which can be tuned by varying 
$\mu$ and $B$,
are given by
\begin{equation}
\label{eq:def-n-m}
n \equiv\frac{1}{N} \sum_{i} \langle  \hat n_{i \uparrow} 
+ \hat n_{i \downarrow}\rangle\,; \quad
m \equiv\frac{1}{N} \sum_{i} \langle\  \hat n_{i \uparrow} 
- \hat n_{i \downarrow}\rangle\,.
\end{equation}
We use in Eq.~(\ref{eq:def-n-m}) and below the notation:
\begin{equation}
\label{eq:def-expet-val}
\Big\langle \hat O\Big\rangle 
\equiv \frac{{\rm Tr}(e^{-\beta \hat H}\, \hat O)}{{\rm Tr} (e^{-\beta \hat H})}\,,
\end{equation}
where $\hat O$ is any operator and $\beta\equiv 1/T$ is inverse temperature.
We measure the density-density correlation 
\begin{equation}
\label{eq:def-C_nn}
c_{nn}(\vec{r}_{ij})= \langle \hat n_i \hat n_j \rangle - \langle \hat n_i \rangle \langle \hat n_j \rangle,  
\end{equation}
and 
singlon-doublon 
correlation 
\begin{equation}
c_{s_{\uparrow}d}(\vec{r}_{ij})
=\langle \hat s_{i_\uparrow} \hat d_j \rangle - \langle \hat s_{i_\uparrow} \rangle \langle  \hat d_j \rangle,
\end{equation}
in both experiment and simulation, where the doublon and (majority-spin) singlon density operators are defined as 
$\hat d_j\equiv \hat n_{j \uparrow} \hat n_{j \downarrow}$ 
and $\hat s_{i\uparrow}\equiv  \hat n_{i\uparrow}-\hat d_i= \hat n_{i\uparrow}-
\hat n_{i \uparrow} \hat n_{i \downarrow}$, 
respectively. 
We use $\Vec{r}_{ij}$ 
to denote the vector connecting site $i$ to site $j$. Under PBC or in the thermodynamic limit, correlation functions only depend on $\vec{r}_{ij}$ 
due to 
translational symmetry of the lattice.
Structure factors can provide useful  
information in Fourier space and 
can be conveniently obtained from 
the correlation functions. For example, the charge structure factor is 
\begin{equation}
S_{\rm cdw}=\frac{1}{N}\sum_{i,j} e^{i \vec{k} \cdot (\vec{r}_i-\vec{r}_j)} c_{nn}(\vec{r}_{ij})\,.
\end{equation}
Additionally, we compute 
the double occupancy,
$D\equiv \langle n_{i\uparrow} n_{i\downarrow} \rangle $ systematically versus $T$.
Spin correlation functions as well as the spin susceptibility
are also computed
\begin{equation}
\label{eq:Sz-suscep}
\chi_{\rm spin}= \frac{1}{N}\sum_{i,j}  \int_{0}^{\beta} \big \langle \hat S^z_{i}(\tau) \hat S^z_{j} (0) \big\rangle d\tau\,, 
\end{equation}
where the integrand is a non-equal imaginary-time correlation
\begin{equation}
\label{eq:def-time-corr}
\big \langle \hat S^z_{i}(\tau) \hat S^z_{j} (0) \big\rangle
\equiv \frac{{\rm Tr}[e^{-(\beta-\tau) \hat H}\, \hat S^z_{i} e^{-\tau \hat H}\, \hat S^z_{j} )}{{\rm Tr} (e^{-\beta \hat H})}\,.
\end{equation}

To probe pairing, we compute 
the
$s$-wave pairing correlation function 
\begin{equation}
P(\vec{r}_{ij})=\langle \hat \Delta_i^\dagger \hat \Delta_j\rangle ,
\end{equation}
where $\hat \Delta_i^\dagger = (\hat c_{i\uparrow}^{\dagger} \hat c_{i\downarrow}^{\dagger}+c_{i\downarrow} \hat c_{i\uparrow})/2$.
The corresponding pairing structure factor is
\begin{equation}
S_{\rm pair}(\vec{k})=\frac{1}{N}\sum_{i,j} e^{i \vec{k} \cdot (\vec{r_i}-\vec{r_j})} \langle \hat \Delta_i^\dagger \hat \Delta_j\rangle\,,
\end{equation}
The 
$s$-wave pair susceptibility is also measured 
\begin{equation}
\chi_{\rm pair} (\Vec{k})= \frac{1}{N}\sum_{i,j} e^{i\vec{k}\cdot (\vec{r}_i-\vec{r}_j)} \int_{0}^{\beta} \langle \hat \Delta_{i}^\dagger (\tau) \hat \Delta_{j}^{} (0) \rangle d\tau,
\end{equation}
using the same definition as
in Eq.~(\ref{eq:def-time-corr}).

A more direct way to characterize pairs with finite momentum is to look at the pair momentum distribution $n_{\vec{Q}}$. We define a pairing matrix for Cooper pairs with center-of-mass momentum $\vec{Q}$  \cite{Rosenberg2015}
\begin{equation}
M_{\vec{k}\vec{k}^\prime,\vec{Q}}=\langle \hat \Delta_{\vec{k}\vec{Q}}^{\dagger} \hat \Delta_{\vec{k}^{\prime}\vec{Q}} \rangle -\delta_{\vec{k}\vec{k}^{\prime}}\langle \hat c_{\vec{k}+\vec{Q} \uparrow}^{\dagger} \hat c_{\vec{k}+\vec{Q} \uparrow}\rangle \langle \hat c_{-\vec{k} \downarrow}^{\dagger} \hat c_{-\vec{k}\downarrow}  \rangle\,,
\end{equation}
with $\hat \Delta_{\vec{k}\vec{Q}}^{\dagger}=\hat c_{\vec{k}+\vec{Q}\uparrow}^{\dagger} \hat c_{-\vec{k} \downarrow}^{\dagger}$. 
We compute the matrix elements 
at each $\vec{Q}$. The matrix is then diagonalized and the leading eigenvalue is denoted as $n_{\vec{Q}}$. 
In the usual unpolarized case, $n_{\vec{Q}=\vec{0}}$ is finite in the thermodynamic limit below the transition temperature~\cite{He2022}, and
gives the condensate fraction.

Comparisons between experiment and computation mostly focus on correlations functions at shorter distances. 
In our computations, we average over all pairs of sites $i$ and $j$ for which $\vec{r}_{ij}=\vec{r}$ under translational symmetry and PBC in computing correlation functions for $\vec{r}$. 
We have performed select computations to confirm that 
the influence of boundary conditions and finite size effects is minimal
for the direct comparisons.  
  
\subsection{Finite-temperature constrained-path AFQMC} 
\label{ssec:FT-AFQMC}
\begin{figure*}[t]
 \includegraphics[width=2\columnwidth]{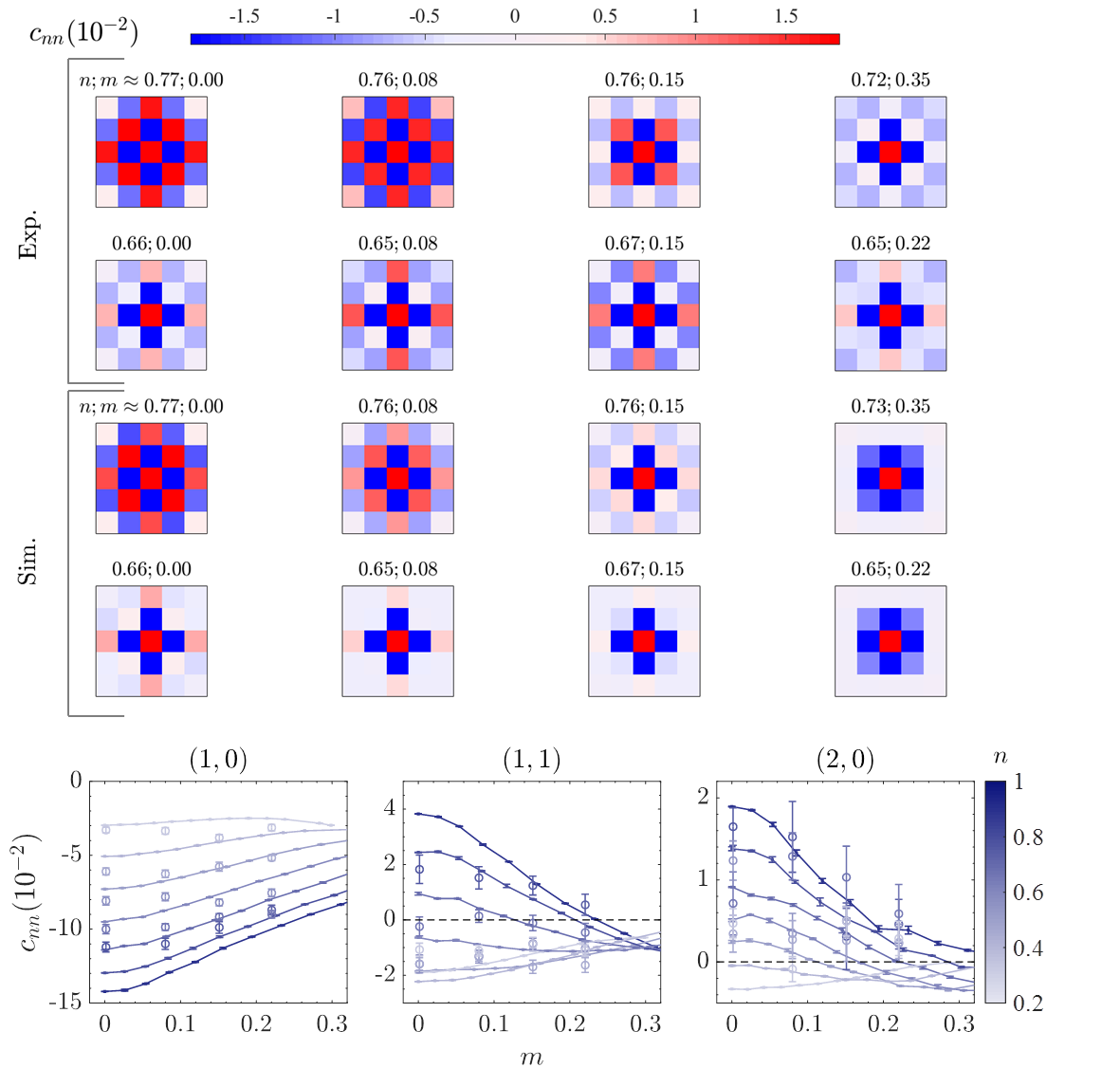}
\caption{\textbf{Comparison between experiment and computation:
density-density correlation. 
}
The upper panel shows color maps of $c_{nn}$,
measured with respect to the site in the middle, for various values of 
density $n$ and magnetization 
$m$ (labeled above each plot as ``$n;m$").
The top half shows experimental measurements, 
while the bottom half shows the corresponding results from AFQMC.
The lower panel shows $c_{nn}$ 
at site  
$\vec{r}=(1,0)$, $(1,1)$, and $(2,0)$, respectively, 
for different densities (indicated by color),
as a function of magnetization. 
Solid lines are from computation and circles are from experimental measurements. 
Results are all for $U \approx -4\,t$ at temperature
$T=1/3\,t$.
}
 \label{fig:Exp_den_den_corr}
 \end{figure*}

Computationally, we employ the finite temperature constrained-path (CP) AFQMC  method \cite{Zhang1999, He2019}.
This state-of-the-art method controls the 
fermion sign problem with a constraint on the path integrals in the space of auxiliary fields, which allows us to perform computations on the Hubbard model in Eq.~(\ref{eq:Hubbard_N1}) systematically for different parameters. When $B=0$, the constraint is not needed and numerically exact calculations 
can be restored as with the standard determinantal quantum Monte Carlo (DQMC) 
approach~\cite{BSS,Scalettar1989,Yuanyao2025}. 
When $B\ne 0$, a sign problem occurs which is removed 
with the CP approximation, and
the method has computational cost which scales the same 
as in sign-problem-free
cases. 

The CP-AFQMC method relies 
on the field-theory formulation to transform a
many-body problem into a superposition of non-interacting problems in 
auxiliary-fields. However, there are two key
reformulations. 
First the partition function is recast as a path integral over a constrained portion of the paths in auxiliary-field space \cite{Zhang1999}. 
The constraint is 
imposed as a gauge condition (sign or phase), based on the projected contribution of a path to the partition function. The constraint is exact when the gauge condition is 
exact \cite{Zhang1999,zhang2019}.  
Second, the path integral is 
performed by branching random walks in imaginary time
which allows
the implementation of the
constraint without any ergodicity difficulties \cite{Zhang1997} that 
would occur under the usual 
Metropolis sampling in DQMC.
In practice, the gauge condition is approximated by 
a trial density matrix using a trial Hamiltonian,  $\hat H_T$. For finite temperatures, 
a single-particle form  
has often been used.
Systematic studies in the 
Hubbard model \cite{He2019,Xiao2023} have been performed using
effective trial Hamiltonians
obtained from
Hartree-Fock (HF) calculations.
Even with a  
$\hat H_T$ from restricted HF the systematic errors
in various equal-time correlation functions are 
within a fraction of a percent compared to DQMC at 
temperatures where the latter can be run \cite{He2019}, while at low $T$
the method smoothly connects to the ground state CP AFQMC \cite{Zhang1997},
where 
many benchmarks 
show that AFQMC is among the most accurate many-body computational methods \cite{LeBlanc2015,Qin2020,Xu2023}.
In this work, a restricted HF $\hat H_T$ is used, with the same parameters as in the target 
many-body $\hat H$, except that chemical potential 
is tuned such that the HF results yield the desired $n$.

\subsection{Quantum gas microscopy} 
\label{ssec:quantum-gas-expt}

 Experimentally, we realize the attractive Hubbard model using a gas of fermionic $^{40}$K atoms in their two lowest hyperfine states, trapped in an optical lattice~\cite{Hartke2023, Hartke2020}. The Hubbard tunneling amplitude $t$ is controlled by the lattice depth, while the interaction strength $U$ can be tuned via the magnetic field.
 A bilayer quantum gas microscope enables the spin-resolved read-out of particle numbers with single-site resolution~\cite{Hartke2023}.
 In an experimental sequence, the Fermi gas is prepared in a single horizontal layer of a 2D optical lattice. For the measurement, the lattice potential is suddenly raised to freeze the atoms' position. Ramping up a vertical superlattice in the presence of a magnetic field gradient transports atoms in a spin-selective manner into an upper (spin up) and lower (spin down) layer, where they can subsequently be imaged sequentially via Raman sideband cooling~\cite{Hartke2023}.
 Repeating the sequence, each time with a newly cooled quantum gas, yields a large sample from which density and magnetization profiles as well as particle correlations can be obtained.
 
 \vskip0.20in \noindent

 \begin{figure*}[t]
 \includegraphics[width=2\columnwidth]{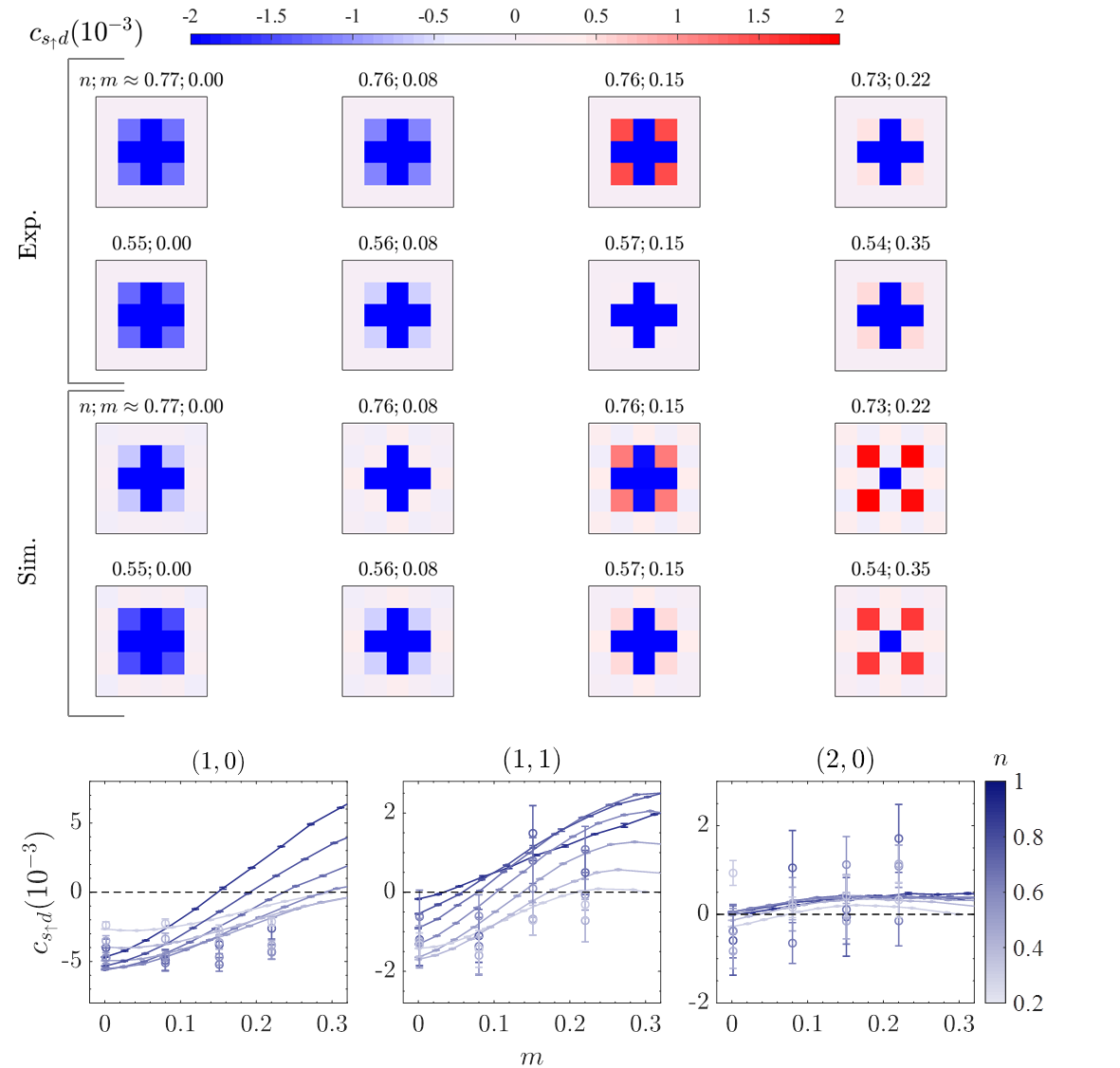}
\caption{\textbf{Comparison between experiment and computation:
singlon-doublon 
correlation.}
The setup is the same as in Fig.~\ref{fig:Exp_den_den_corr},
but now for $c_{s_{\uparrow}d}({\vec r})$. 
In the color maps, $c_{s_{\uparrow}d}$ values statistically indistinguishable from zero are plotted as zero for clarity. 
}
 \label{fig:Exp_singlonup_doublon_corr}
 \end{figure*}
\section{Results} 
\label{sec:Results}

Here we present our 
results in three parts.
In Sec.~\ref{sec:hand-shake}, we perform detailed comparisons between experiment and computation. 
Over 
a wide range of density and magnetization, excellent agreement is seen between direct quantum gas microscopy 
measurements and AFQMC computation.  
In Sec.~\ref{ssec:temperature dependence} we use AFQMC 
to systematically investigate the spin-imbalanced Hubbard model down to low temperatures.
The temperature evolution of various properties 
are characterized, 
including FFLO correlations and other signatures.
Finally in Sec.~\ref{ssec:Phase diagram of FFLO order and its detection}, we 
summarize our data across the desity-magnetization plane in the form of ``phase diagrams'' in terms of the presence of FFLO correlations. A path towards experimental observation of FFLO is then outlined. 
 
\subsection{Density and magnetic correlations -- handshake between computation and experiment}
\label{sec:hand-shake}

\begin{figure*}[t]
 \includegraphics[width=2\columnwidth]{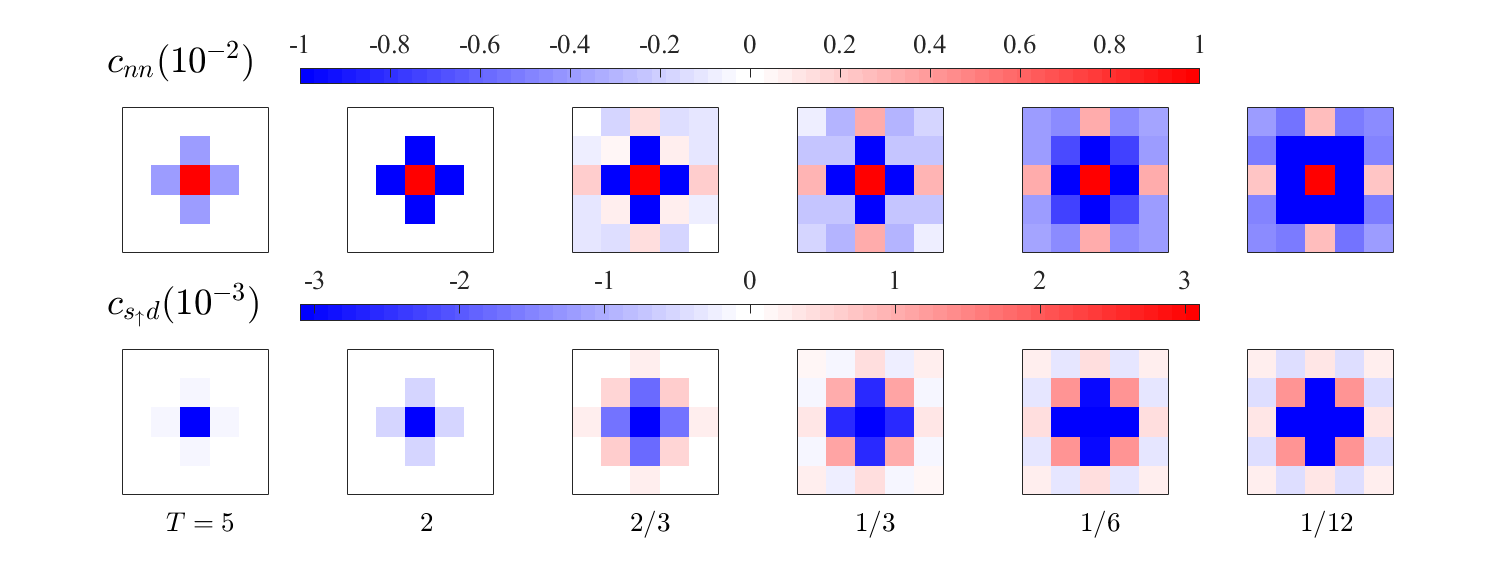}
\caption{\textbf{Temperature evolution of 
short-range density-density 
and singlon-up-doublon 
correlations.} The color represents the correlation magnitude, which is  measured with respect to the site in the middle. 
We show the correlation for ${\vec r}$ up to two sites away in each dimension. The calculations are performed on a $16 \times 16$ square lattice with density $n \approx 0.7$ and magnetization $m \approx 0.15$ at $U=-4$.
}
 \label{fig:den_den_sing_doub_vs_T_n07_m015}
 \end{figure*}

We first
make direct and systematic comparisons between experiment and simulation 
on short-range charge and spin correlation functions.  
The density-density correlations $c_{nn}$ are 
shown in Fig.~\ref{fig:Exp_den_den_corr} for a range of densities $n$ and magnetizations $m$
obtained from both experiment and AFQMC simulations at $U=-4 t$ and $T = 1/3 t$.The correlation $c_{nn}({\vec r})$ is measured with respect to the site in the middle, for a range of separations ${\vec r}$.
The correlation patterns are shown only for a $5\times 5$ central patch, although the actual measurements are on a bigger lattice in both experiment and computation, as discussed in Sec.~\ref{sec:System and Overview of Methodology}.
We see that the qualitative features  
seen in experiment and 
simulation are largely consistent, also considering uncertainties in the thermometry in the experiment ($\sim\pm 0.1 t$~\cite{Hartke2020}).

At a relatively high density $ n \approx 0.76 $ as shown in the top rows for both experiment and simulation, 
significant short-range charge-density wave (CDW) correlation is present
in the spin-balanced system.
With increasing magnetization, 
as we move to the right across each row, the 
CDW pattern quickly shrinks.
As the density is decreased to $n \approx 0.66$ 
(as shown in the second row for both experiment and simulation),
the CDW correlation becomes 
considerably weaker. 
The contrast with $ n \approx 0.76 $ is 
especially strong at low magnetization. 
As the magnetization increases,
the pattern
becomes less different
between the two densities;
at the largest 
$m$ values shown,  
CDW correlations are no longer present: $c_{nn}$ is now negative rather than positive
at the closest diagonal site ${\vec r}=(1,1)$. 
Interestingly, in the doped repulsive Hubbard model with density $n \sim 0.7$, spin correlation at $(1,1)$ also changes sign first when the system enters the stripe phase~\cite{Feng2025}.
The sign change in the charge correlation here 
is thus likely a precursor to the appearance of FFLO order, which we explore systematically below. 

A change of character is seen in 
the charge correlation at higher densities, as the 
magnetization is varied.
This is most visible in the $(1,1)$ and $(2,0)$ 
correlations  in the bottom panel, which for sufficiently large $n$ change sign
as a function of $m$.
At high densities $c_{nn}({\vec r})$ is
positive at both of these ${\vec r}$ values 
near spin balance, because of the 
CDW correlation. As $m$ increases, 
the system turns into a polarized fluid.
In the limit $m\rightarrow n$, 
$c_{nn}$ reflects the exchange hole of the majority spin fermions, which is short-ranged and negative.  
The sign change results from balancing  these two competing tendencies as $m$ is increased.

In the context of possible FFLO order, an important question is the spatial  distribution of the excess fermions.
In Fig.~\ref{fig:Exp_singlonup_doublon_corr} we show the 
singlon-up-doublon correlation function $c_{s_{\uparrow}d}$
in a similar layout to 
Fig.~\ref{fig:Exp_den_den_corr}. Good agreement is again seen between experiment and simulation. 
In this parameter regime, 
the singlon-doublon correlation is considerably weaker than the charge-charge correlation, and rather short-ranged. 
At small magnetization, $c_{s_{\uparrow}d}(1,0)$
is negative, reflecting the
Pauli exclusion between the singlon  
and the like-spin fermion in the doublon.  
The same trend is also exhibited in 
$c_{s_{\uparrow}d}(1,1)$,
especially at lower densities. 
As magnetization $m$ is increased, 
a competing tendency appears, 
and the short-range $c_{s_{\uparrow}d}$ correlation turns positive. 
In the  
limit with a single $\downarrow$-fermion, the presence of other $\uparrow$-fermions can increase the mobility of the doublon,   and effectively lower kinetic energy. This is manifested as a change of sign in the short-range singlon-doublon correlations, 
particularly visible at higher density in $c_{s_{\uparrow}d}(1,0)$ and in a wider density range 
in $c_{s_{\uparrow}d}(1,1)$.

\subsection{Temperature dependence and evolution of FFLO correlations}
\label{ssec:temperature dependence}

 \begin{figure*}[t]
 \includegraphics[width=2\columnwidth]{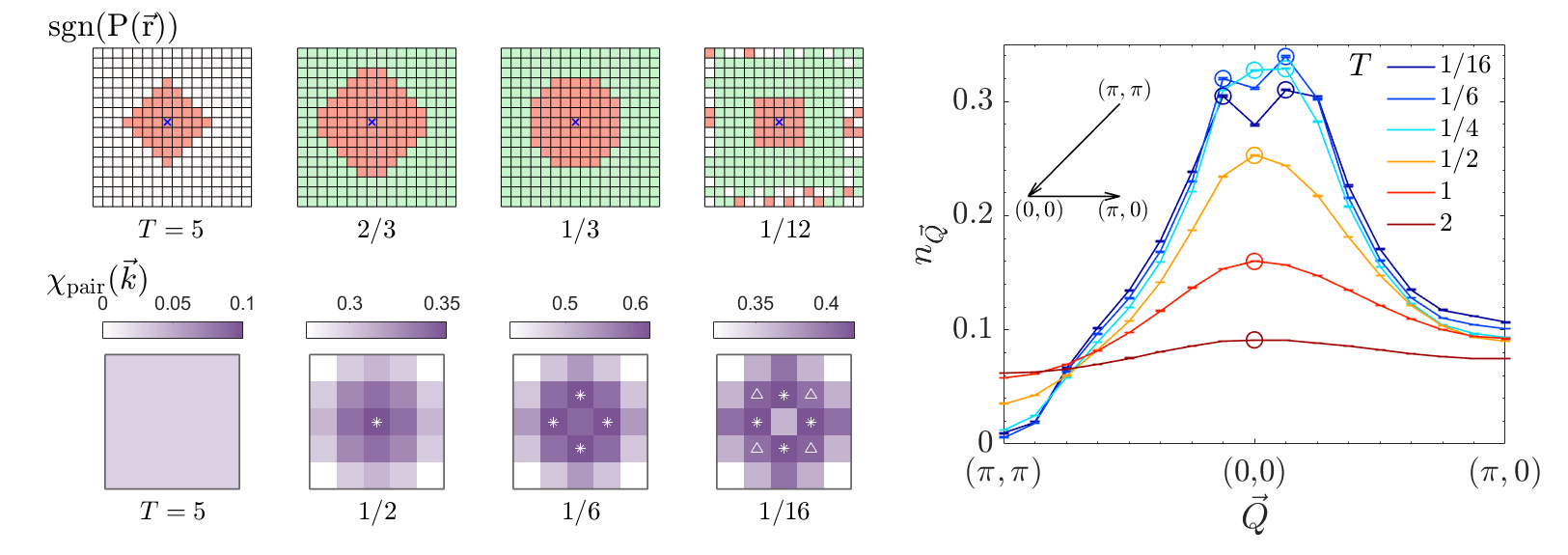}
\caption{\textbf{
Appearance of FFLO order as a function of 
temperature.} 
Top left:
Sign of 
$s$-wave pairing correlation in real space as a function of temperature.
The reference point is 
in the middle of the lattice, denoted by ``x". Orange represents positive pairing correlations while green represents negative. 
As $T$ decreases,
the pattern evolves from a diamond to a circle and then to 
a square.
Bottom left:
$s$-wave pairing susceptibility $\chi_{\rm pair}(\vec{k})$ 
in the vicinity of $\vec{k}=(0,0)$, versus temperature.
As $T$ lowers,
the position of the peak changes from 
$(0,0)$ to $(\frac{\pi}{8},0)$, 
and then also spreads to 
$(\frac{\pi}{8},\frac{\pi}{8})$,
reflecting the 
trend of $P(\vec{r})$
in real space. 
The peak positions are indicated by $*$
(and also $\triangle$ for comparable peak 
values at $T=1/16$).
Right:
Pair momentum distribution $n_{\vec{Q}}$ vs.~$\vec{Q}$,
plotted along linecuts 
(see inset) for six temperatures. 
Circles
mark the positions of the peak, 
which shift
away from 
$|\vec{Q}|=0$ as $T$ is lowered.
The system is a $16 \times 16$ periodic square lattice with density $n \approx 0.7$ and magnetization $m \approx 0.15$.
}
 \label{fig:signals_of_FFLO}
 \end{figure*}

In order to  understand the relation between the correlations measured above at $T=1/3$  and the properties 
at low temperatures,
we next probe the temperature dependence of the system. 
We first examine 
the evolution of the two correlation functions discussed in Sec.~\ref{sec:hand-shake}, 
$c_{nn}$ and $c_{s_{\uparrow}d}$. 
In Fig.~\ref{fig:den_den_sing_doub_vs_T_n07_m015}, these correlations are shown 
as a function of temperature,
for a typical system with $n \approx 0.7$ and $m \approx 0.15$. We see that the short-range correlation patterns at $T=1/3$ grow in their range as the temperature is lowered, but the evolution is smooth and continuous. 
Interestingly, the singlon-doublon correlation here develops a checkboard pattern at low $T$,
suggesting a singlon (excess spin) density wave; 
in contrast, a hint of CDW correlation appears in $c_{nn}$ at elevated temperature ($T=2/3$)
before it fades away as $T$ is lowered.

 \begin{figure}[t]
 \includegraphics[width=1\columnwidth]{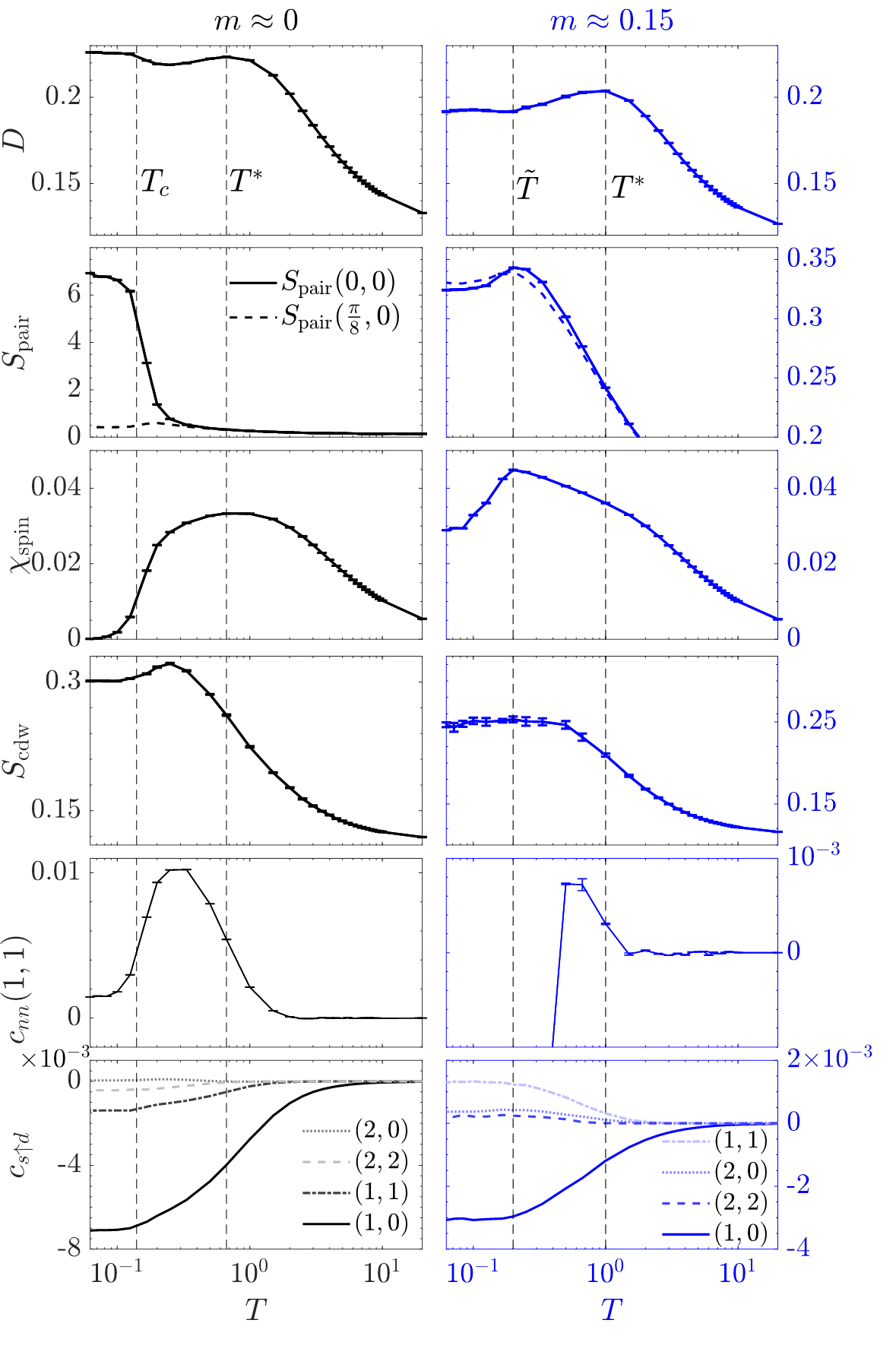}
\caption{\textbf{
Properties and temperature-dependence in spin-balanced vs.~spin-imbalanced systems.} 
Different 
correlation functions and related quantities are 
shown 
as a function of temperature $T$, for $m=0$ (left column) and $m = 0.15$ (right column), 
with otherwise the same parameters. 
$T^{*}$ denotes the first peak position in the double occupancy $D$ as $T$ is lowered; $T_c$ in $m=0$ 
is the BKT superfluid transition temperature \cite{Paiva2004}; and $\Tilde{T}$ 
in $m=0.15$ denotes the temperature below which the peak in the pairing structure factor shifts to a finite momentum. 
 The vertical dashed lines are 
 a guide to the eye.
The system has $U=-4$ and
density $n \approx 0.7$. All calculations are performed on a $16 \times 16$ lattice.}
\label{fig:Fig6_D_Spair_Chispin_Scdw_singdoub_vs_T}
\end{figure}

We now turn to the question of 
FFLO correlations, which we characterize with 
three different quantities:
(a) real-space pairing correlations $P(\vec{r})$, (b) pairing susceptibility $\chi_{\rm pair}(\vec{k})$, and (c) pair momentum distribution $n_{\vec{Q}}$.
These are defined in Sec.~\ref{sec:System and Overview of Methodology},
and can be computed straightforwardly in CP-AFQMC
after the auxiliary-field paths have been sampled.
The results are summarized in 
Fig.~\ref{fig:signals_of_FFLO}
for the same system 
as in Fig.~\ref{fig:den_den_sing_doub_vs_T_n07_m015}.
They show a consistent picture of the appearance 
of FFLO correlations when $T$ is lowered,
as we discuss in further detail below.

The sign of the pair correlation function $P(\vec{r})$ is shown in $\vec{r}$-space in
the entire simulation cell,
for four $T$ values
as indicated under each plot. 
A site is left blank (gray) if the correlation magnitude is smaller than $10^{-12}$ or within the statistical error bar. At high temperature, 
the correlation decays exponentially and only near neighbor correlations are visible. As the temperature is lowered, 
the correlation range grows and $P(\vec{r})$ turns negative beyond some distance, indicating the appearance of a first node in the pairing correlation. The real space pattern of pairing correlation 
evolves from a diamond at high $T$ 
to a circle at $T=1/3$, and finally to a square when the
temperature is further lowered. At $T=1/12$, some of the correlations near the edges become positive again, 
indicating the appearance of the second node. 

The pairing susceptibility $\chi_{\rm pair} (\vec{k})$ is also displayed 
in Fig.~\ref{fig:signals_of_FFLO},  
for a
region 
in the vicinity of 
$\vec{k}=(0,0)$.
The momentum-space resolution 
in this system is $\pi/8$,
represented by a unit square.
At $T=5$, the pair susceptibility is almost independent of $\vec{k}$, 
as the pairing correlation
is dominated by the on-site term and decays rapidly with distance.
As $T$ decreases to $T=1/2$, 
a peak appears at $(0,0)$ (although in the 
real-space  
pairing correlation,  
negative values are already present 
at this temperature). When $T$  
is lowered to $1/6$, the peak splits into four and moves
to non-zero momenta $\vec{k}=(\pi/8,0)$ and its symmetry-related values.
Upon further cooling, 
another set of peaks is observed along the diagonals
$(\pi/8,\pi/8)$ etc.
The evolution in the peak locations of $\chi_{\rm pair} (\vec{k})$
is connected with the changes in the
pattern of the real-space
pairing correlation function 
$P(\vec{r})$
discussed above. 
The lack of a precise correspondence (e.g.~appearing at exactly the same $\vec{k}$ point) is due to the details of the
correlation functions and finite lattice size effects. 
(Competition between the FFLO wave vector at $(1,0)$ and $(1,1)$  has also been 
studied
in Hartree-Fock-Bogoliubov calculations for the ground state 
\cite{Chiesa2013}.)

The third quantity, the pair momentum distribution, 
provides a direct measure which  
further corroborates the observations above. In 
Fig.~\ref{fig:signals_of_FFLO}
the pair momentum distribution $n_{\vec{Q}}$ is plotted
at six different temperatures,
as a function of $\vec{Q}$ along a path in $\vec{Q}$-space. 
As temperature decreases,  $n_{\vec{Q}}$ 
starts to develop a peak 
at $\vec{Q}=(0,0)$, which grows 
more pronounced at lower $T$,
reflecting enhanced pairing correlations.
By $T=1/6$, 
the peak has shifted away to non-zero momentum, 
signifying the emergence of 
FFLO correlations.
At the lowest temperature $T=1/16$, the peak value is seen to decrease slightly compared to 
$T=1/6$. 
This is due to finite lattice size effects: 
at low $T$ the peak location falls 
between $(\frac{\pi}{8},0)$ and $(\frac{\pi}{4},0)$, 
incommensurate with the $16 \times 16$ lattice here (see Supplementary Material).

Having established the 
key signatures for FFLO in these systems, 
we now
provide further, systematic quantifications of 
the properties and their
temperature progression.
We do so by contrasting
the 
spin-imbalanced system
($m\ne 0$)
with the spin-balanced system 
($m= 0$), whose properties are better understood.
In Fig.~\ref{fig:Fig6_D_Spair_Chispin_Scdw_singdoub_vs_T}
we study a variety of quantities as a function of temperature, 
comparing the two cases side by side. The quantities, shown as successive rows for each system, include:
double occupancy (doublon density) 
$D = \langle \sum_i \hat d_i\rangle/N$,
the $s$-wave pairing structure factor $S_{\rm pair}$, spin susceptibility
$\chi_{\rm spin}$, CDW structure factor $S_{\rm cdw}$, density-density correlation $c_{nn}$, and singlon-up-doublon correlation $c_{s_{\uparrow}d}$,
all as defined in Sec.~\ref{sec:System and Overview of Methodology}.

We first consider the double occupancy $D$.  
The spin-balanced system shows an interesting non-monotonic behavior as a function of $T$.
At high temperatures, 
it approaches the non-interacting limit 
$D \rightarrow n_{\uparrow}  n_{\downarrow} =n^2/4$, which is $\sim 0.123$ for the system 
in Fig.~\ref{fig:den_den_sing_doub_vs_T_n07_m015}.
As $T$ decreases, $D$ increases due to correlation effects from the attractive on-site interaction. 
At a temperature we have denoted 
$T^\star$, 
which signals the appearance of preformed pairs, 
the system reaches a local maximum in $D$
(entering what is sometimes referred to as the 
pseudogap regime), after which $D$ 
decreases. 
This decrease is presumably the consequence of enhanced exchange as pairs gain coherence. 
The curve eventually turns and rises again as $T$ is further reduced to the superfluid transition temperature $T_c$, and $D$
increases to approach its ground state value.  
The behavior of $D$ mirrors that seen in the repulsive 
Hubbard model at half-filling and weak interaction~\cite{Schafer2021,Song2025B}, with the trends inverted due to particle-hole symmetry. A more detailed analysis of the behavior of $D$ in the repulsive model at half-filling is given in 
Ref.~\cite{Schafer2021,Song2025B}. (Note that the system there is spin-balanced while
our system at $n=0.7$ and $m=0$ corresponds to 
a spin-imbalanced situation in the repulsive model.) Now turning to the spin-imbalanced system on the right, we see that the $D$ vs.~$T$ curve follows a very similar trend. The double occupancy of the spin-imbalanced system is always lower than the spin-balanced case. At high $T$, 
its asymptotic value is $D\rightarrow (n^2-m^2)/4 \sim 0.1$, and with lowering $T$ it is  
bounded by the density of the minority spin, $(n-m)/2$.
As in the spin-balanced case, we have
labeled the location of the first maximum of 
$D$ as $T^\star$.  
Below $T^\star$, the double occupancy in $m=0.15$ decreases as for $m=0$, 
but it does not display as clear and pronounced a rise again when $T$ is further lowered. 
This has to do with the appearance of 
pairs with finite momentum
as we discuss next.

 \begin{figure*}[t]
 \includegraphics[width=2\columnwidth]{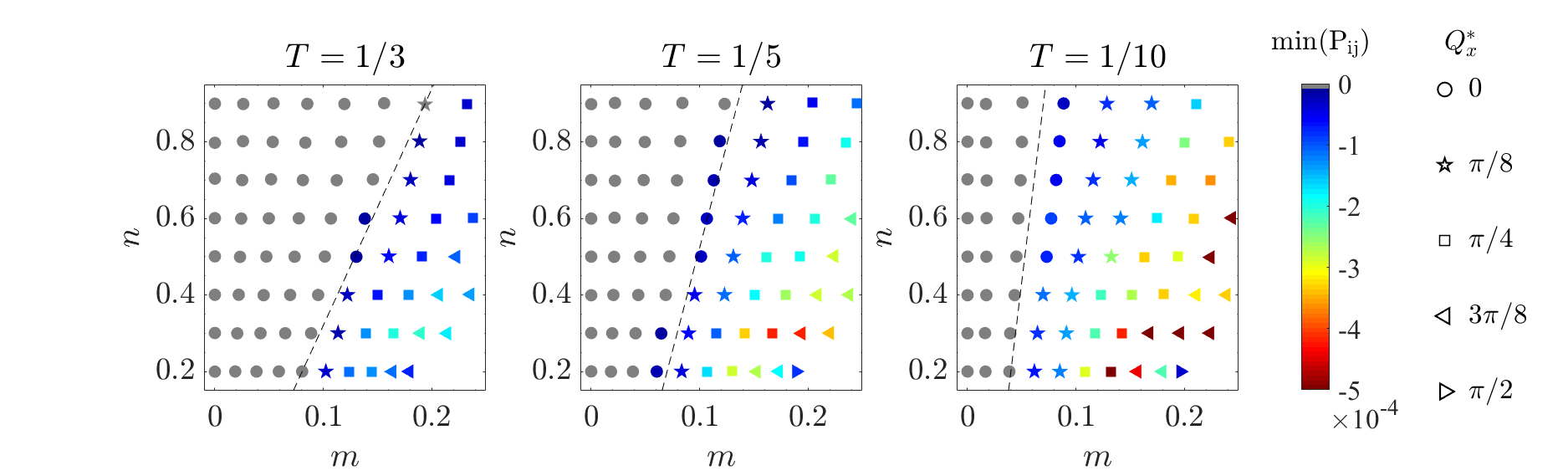}
\caption{\textbf{
Presence of 
FFLO correlations  
at three 
temperatures: $T=1/3, 1/5$, and $1/10$.} 
Two signatures to detect FFLO order, 
pair momentum distribution $n_{\vec{Q}}$ and 
negative values in the pairing correlation $P(\vec r)$, are scanned
as a function of density $n$ and magnetization $m$.
 Symbol shapes represent different peak positions $(Q_x^*,0)$ 
of
$n_{\vec{Q}}$, as indicated by the legends on the right. A non-zero position (symbols other than a circle) indicate pairs of non-zero momentum.
Symbol color represents the most negative value of $P(\vec r)$.
($P(\vec r)>0$ for all $\vec r$ is recorded as 0 in the plot.) The dashed line is a guide to the eye, separating normal pairing (grey) from presence of FFLO correlations 
(other colors).
Calculations are performed on a $16 \times 16$ square lattice, with $U=-4$.
}
 \label{fig:phase_diagram}
 \end{figure*}

The $s$-wave pairing structure factor $S_{\rm pair}(\vec{k})$ develops a sharp peak at $\vec{k}=(0,0)$ as the $m=0$
system undergoes a Berezinskii–Kosterlitz–Thouless (BKT) phase transition 
at $T_c$ ($T_c \sim 0.14$ \cite{Paiva2004} in the system in Fig.~\ref{fig:Fig6_D_Spair_Chispin_Scdw_singdoub_vs_T}), as shown in the second row in the figure. 
In contrast $S_{\rm pair} (0,0)$ for the spin-imbalanced  case is much smaller;
as $T$ is lowered, it is exceeded by $S_{\rm pair} (\pi/8,0)$ at a temperature
$\Tilde{T}$, indicating 
increasing presence of 
pairs with finite momentum. 
$S_{\rm pair}(\pi/8,0)$  decreases slightly at lower $T$ due to incommensurate finite-size effects, as seen above in $n_{\vec{Q}}$ (Fig.~\ref{fig:signals_of_FFLO}). 
We use $\Tilde{T}$ as a proxy for the appearance of FFLO order.
It should be emphasized that  
$\Tilde{T}$ is not a BKT transition temperature as $T_c$ is in the case of $m=0$, but rather a kind of upper bound.
As seen in the discussions
in Fig.~\ref{fig:signals_of_FFLO}, $\Tilde{T}$ is more a signal for the appearance of FFLO correlations.

The spin susceptibility $\chi_{\rm spin}$ is shown next. 
In the spin-balanced case, 
the peak position of $\chi_{\rm spin}$ is sometimes used to define $T^\star$,
the pseudogap temperature \cite{Paiva2004} below which pairs start to form. We see that this coincides exactly 
with the first maximum in $D$ as discussed earlier. 
As $T$ is lowered, 
there is an abrupt decrease in the spin susceptibility around $T_c$, below which $\chi_{\rm spin}$ almost vanishes, reflecting the superfluid pairing gap. 
For the spin-imbalanced system,  we see that $\chi_{\rm spin}$ behaves similarly to the 
$m=0$
case when $T > T^* $. As $T$ decreases, $\chi_{\rm spin}$ begins to differ significantly 
and continues to grow to a maximum at $\tilde{T}$, where pairs with finite momentum develop. Below $\tilde{T}$, $\chi_{\rm spin}$ decreases but remains finite in the low-temperature limit due to the spin fluctuations of the excess fermions.
 
Charge correlations are examined in the next two rows in
Fig.~\ref{fig:Fig6_D_Spair_Chispin_Scdw_singdoub_vs_T}.
In the spin-balanced case, 
a peak emerges in $S_{cdw}$ in the pseudogap regime, consistent with the picture of pairs gaining coherence 
to form a short-range CDW spatial pattern. As pairs become more coherent (and with increasing $D$), the CDW order 
decreases, as is particularly evident in the density-density correlation $c_{nn}(1,1)$. 
In the spin-imbalanced system, 
$S_{\rm cdw}$ is smaller and there is not a clear peak at intermediate temperatures. 
The magnitude of $c_{nn}(1,1)$ 
is $\sim 10$ times smaller;   
furthermore, it 
changes sign rapidly after the formation of a weaker CDW pattern at $T \sim 0.35$, which is
illustrated by the pattern 
in Fig.~\ref{fig:den_den_sing_doub_vs_T_n07_m015}. 

Finally the singlon-doublon correlations are analyzed in the last row.
In the spin-balanced case, 
$c_{s_{\uparrow}d}(\vec{r})$ is negative for all the short-range $\vec r$ displayed. Even at high $T$,
a weak repulsion is seen between a doublon and a singlon. The repulsion, driven by the exchange hole between the singlon and the like spin in the doublon, strengthens considerably below $T^\star$. 
In the spin-imbalanced case, the nearest-neighbor $c_{s_{\uparrow}d}(1,0)$ remains 
almost as negative. However, 
$c_{s_{\uparrow}d}(\vec{r})$ turns (slightly) positive at greater distances,
particularly for $\vec{r}=(1,1)$
as is evident from Fig.~\ref{fig:den_den_sing_doub_vs_T_n07_m015}. The 
effective attraction between an excess spin and a doublon is driven 
by kinetic energy gain. 
The correlation 
starts to develop at $T^*$ and becomes flattened at $\tilde{T}$ with the formation of finite-momentum pairs.

\subsection{Phase diagram of FFLO order and its detection}
\label{ssec:Phase diagram of FFLO order and its detection}

We can now map out a ``phase diagram'' of FFLO correlations with systematic computations in the 
$m$-$n$ plane. The result is shown in Fig.~\ref{fig:phase_diagram} for three temperatures: $T=1/3$, $1/5$, and $1/10$. As demonstrated in Fig.~\ref{fig:signals_of_FFLO}, 
several different probes can be used to detect FFLO correlations, which all yield consistent results, with the caveats of somewhat different quantitative influence by  finite-size effects. Here we show the results from two of these probes together: the $s$-wave pairing correlation function in real space, 
$P(\vec{r})$, and the pair momentum distribution function, $n_{\vec{Q}}$. 
In Fig.~\ref{fig:phase_diagram}, 
the most negative value of 
$P(\vec{r})$ is represented by the color. If the pairing correlation is always positive for any $\vec r$ (as is the case when $m=0$), 
it is set to $0$. Thus, grey stands for the normal superfluid state while other colors represent the 
the appearance of FFLO correlations.
Similarly, if the
pair momentum distribution $n_{\vec{Q}}$ has a peak position $\vec{Q}\ne \vec{0}$, it signals a correlation different from normal 
pairing. As seen in the figure, the ``phase'' boundaries given by these two probes are consistent with each other. It is important to note,
as mentioned in Sec.~\ref{ssec:temperature dependence} in the discussion of $\tilde T$ (spin-imbalanced) versus $T_c$ (spin-balanced), the boundaries here 
do not imply locations of phase transitions. Rather they signal 
the appearance of FFLO correlations of significant range and provide a gauge of their characters.

There is a number of striking features in the result in Fig.~\ref{fig:phase_diagram}. 
Significant FFLO correlations are seen at $T=1/3$, which is well within reach of current experiment.
These correlations are short-ranged but 
sufficient to signal unconventional pairing; they exist in a broad range of density and magnetization,
in which current experimental parameters fall comfortably. 
At a fixed density, it appears that 
the FFLO signals are more visible as the magnetization increases: 
the minimum of the pairing correlation grows more negative and 
the peak of the pair momentum distribution shifts further away from $(0,0)$. 
In ground-state calculations, long-range FFLO order appears rather fragile, and they seem easier to detect at high densities (optical lattice setting) versus low densities (towards the continuum Fermi gas) and at smaller magnetizations \cite{Vitali2022}. How to 
reconcile these observations in a more 
quantitative manner remains an open question.
It is clear, however, that 
the exotic pairing states at low $T$ are natural evolutions from $T=1/3$. 
As temperature is reduced, the region of FFLO order expands, and the signal grows. 
The FFLO wave vector depends on $m$ and $n$. 
We find that in many of the cases studied here, the relation $\left| q \right| =m\pi$ holds at least approximately (see SM).
(Mean-field calculations show \cite{Chiesa2013} that in the ground state this relation holds precisely  at high densities,
while for other parameter regimes the order is incommensurate 
and the wave vector dependence is more complicated.) 

\begin{figure}[h]
 \includegraphics[width=1\columnwidth]{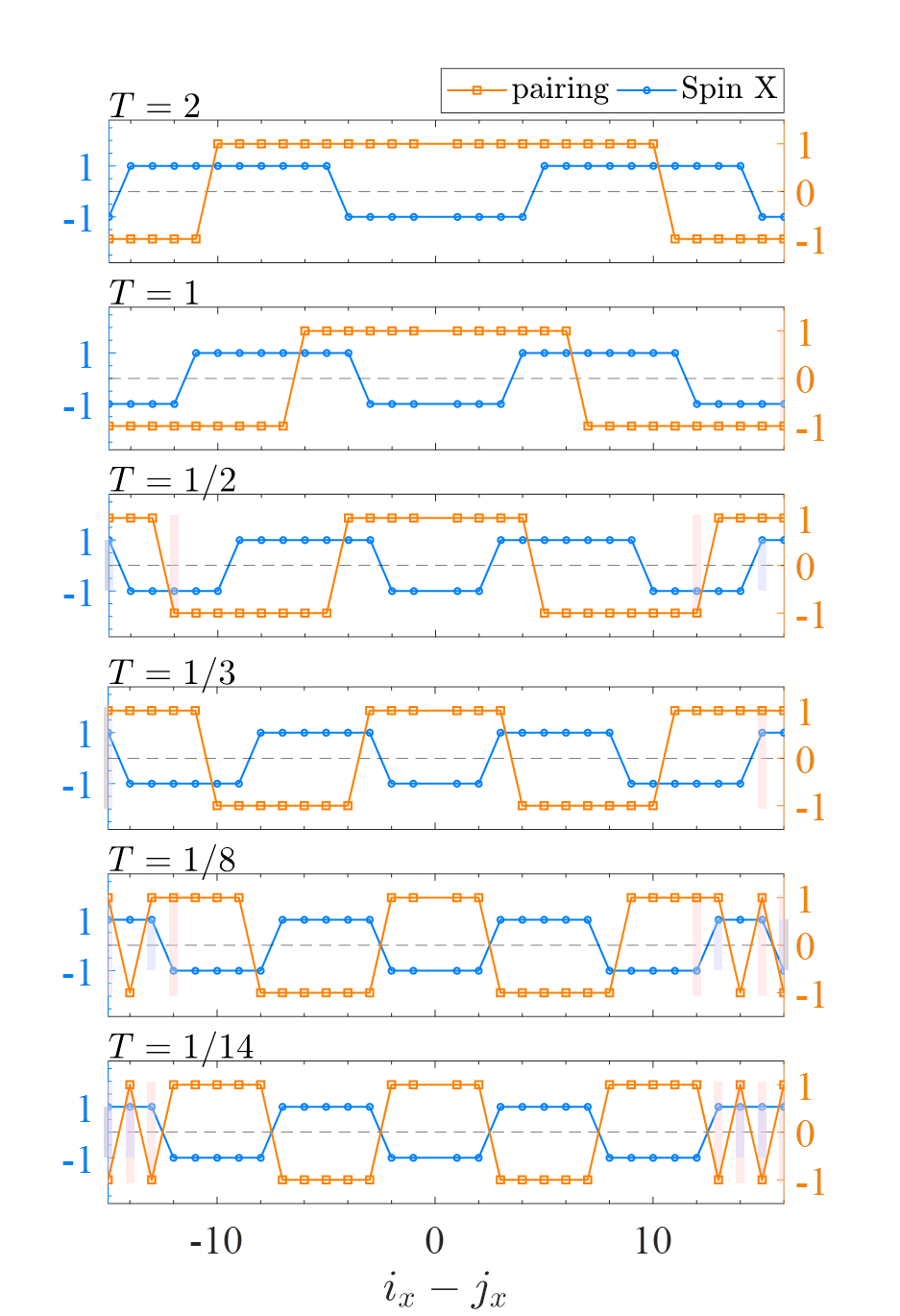}
\caption{\textbf{In-plane spin-spin correlation as a proxy to detect FFLO order.}
The pairing and spin correlation functions in the $x$-$y$ plane
are plotted for a sequence of temperatures, along a linecut in the $x$-direction for a $64 \times 4$ lattice with $n \approx 0.7$ and $m \approx 0.15$.
The reference point is at the origin and the onsite term is omitted.
For clarity, only the signs are plotted, and the two correlations are shown on different scales (left and right vertical axes). 
At large distance the statistical 
error bars become comparable to the signal, indicated by
the light color shaded regions. 
}
 \label{fig:pairing_spin_X}
 \end{figure}

Experimentally it is challenging at present to directly measure pairing correlation functions, or the other signatures we have used in the computations above. Given the exquisite capability of quantum gas microscopy to measure various magnetic and charge correlations, it is highly desirable to seek proxies for detecting FFLO correlations. 
As we have shown, there exists interesting and sometimes subtle interplay between the FFLO correlations and both the 
charge- and  
$z$-direction spin correlations 
(including singlon and doublon
correlations of various forms). 
However, we have not found a direct connection between their behavior and FFLO correlations which applies across different parameter regimes.

Instead we find that 
the spin-spin correlation measured 
in the $x$-$y$ directions, 
$ \langle \hat S^{xy}_{i} \hat S^{xy}_{j} \rangle$,
can potentially serve as a proxy for 
the FFLO order, as  illustrated in Fig.~\ref{fig:pairing_spin_X}.
One way to motivate this result is through the relation between 
the current system and the repulsive Hubbard model, which are related by a
partial particle-hole transformation \cite{scalettar2016, Xie2025, Feng2020} as mentioned earlier. 
The doped, spin-balanced 
repulsive Hubbard model features a
spin-density wave or stripe phases \cite{Chang2010,Zheng2017,Xiao2023}, where the modulated AFM order is  often accompanied by a charge density modulation. Mapped to the attractive Hubbard model, this corresponds to $n=1$ with modest $m>0$, and it would imply a CDW accompanied by modulation in $\hat S^{z}$ AFM correlations. 
In the repulsive Hubbard model
a degenerate spin order is also present in the $x$- or $y$-direction, which translates into superfluid order in the attractive model.
When the attractive Hubbard model is doped, this superfluid order dominates the CDW order. 
Correspondingly, in our present case with spin imbalance, it is reasonable to expect an FFLO phase (with modulation in the pairing correlations) to prevail compared to the CDW. 
The  $ \hat S^{xy}$ correlations in the attractive model can be thought of as the partner channels to pairing, in the sense of $  \hat S^{z} $ being partner to the CDW
(mirroring the relation between charge to spin in the stripe state in the repulsive model).

In Fig.~\ref{fig:pairing_spin_X} we show the temperature evolution of the FFLO and 
$ \langle \hat S^{xy}_{i} \hat S^{xy}_{j} \rangle$ spin correlations. To quantify the modulation wavelength dependence, the calculations are performed in a rectangular  $64 \times 4$ simulation cell with PBC in both directions.
The reference site is at the origin and the correlations are plotted 
along a line cut, with site distance in $y$-direction fixed at $i_y-j_y=0$. 
We see that
the $ \hat S^{xy}$ correlation
shows a modulation similar to  pairing, and the two 
are well synchronized 
(although the spin correlation is weaker, for example by roughly a factor of $4$ near the position of the first node). 
At high temperatures, 
the spin correlation
has a somewhat smaller modulation wavelength than the FFLO pattern.
When temperature 
is lowered, the wavelengths of both the spin $x$-$y$ and FFLO correlations decrease. Finally when $T$ is sufficiently low,
at $T\sim 1/8$, the two patterns mirror each other, with a modulation wavelength of $\lambda \sim 10$.
Interestingly, the modulation wavelength of the $ \hat S^{xy}$ correlation seems much less sensitive to $T$, and reaches this value at a rather high temperature of $T \sim 1/2$, well within reach of current experiments. 
To measure spin $x$ and $y$ correlations in the Hubbard model using cold atoms, it is necessary to access the transverse spin components $\hat{S}^x$ and $\hat{S}^y$. This requires a combination of spin rotations and site-resolved imaging techniques. Before readout, the spin states of the atoms are rotated into the $z$-basis using radio-frequency 
or microwave pulses. For example, a $\pi/2$ pulse about the $y$-axis of the Bloch sphere maps the $\hat{S}^x$ component onto the $\hat{S}^z$ axis, making it measurable via standard spin-resolved imaging. Similarly, a $\pi/2$ pulse about the $x$-axis allows measurement of the $\hat{S}^y$ component. Once the spin rotation is performed, repeated projective measurements of the system are conducted to collect spin population statistics. These measurements are used to calculate the spin correlation functions, such as $\langle S^x_i S^x_j \rangle$ and $\langle S^y_i S^y_j \rangle$, which provide insights into the connection between $\hat{S}^{xy}$ and pairing correlations. We consider this an intriguing direction for future investigation.

\vskip0.20in \noindent
\section{Summary and Conclusion} 
\label{sec:Discussion}

We have employed a combination of computational and experimental approaches to study the 2D doped, spin-imbalanced 
attractive Hubbard model, focusing on understanding and realizing 
exotic superconductivity and pair-density wave states. Using state-of-the-art CP-AFQMC and quantum gas microscopy, we systematically investigate the magnetic, charge and pairing correlations and their temperature evolutions. A quantitative handshake is demonstrated  
at high and medium temperatures between quantum gas microscopy measurements by an experimental emulator --- using $^{40}$K atoms in an optical lattice --- and computation by AFQMC.  
Detailed results are then obtained 
with computation down to low temperatures on 
a variety of observables and correlation functions,
characterizing  the temperature dependence  systematically.
Signatures for FFLO pairing are found to appear at
rather 
high temperatures which evolve smoothly as temperature is lowered and longer-range correlations develop. 
A ``phase diagram'' on the character of these correlations is obtained at several temperatures for a wide range of densities and magnetizations. 
We find that the spin-spin correlation in the $xy$-plane can potentially serve as a proxy for 
detecting FFLO order with the current experimental setup. 

Our work presents a new 
paradigm for understanding and characterizing 
exotic superconductivity. The combination of many-body computation and atomic gas microscopy 
creates a unique and powerful synergy. Recent experimental 
progress has realized a variety of important models 
using optical lattices \cite{Hartke2023,MuqingXu2023,Lebrat2024,Xu2025,Chalopin2024,Mongkolkiattichai2023,Guardado-Sanchez2020,Prichard2024,Bourgund2025,Bakr2025,Bloch2008, Giorgini2008,Mitra2018,Gall2020,Chan2020,Brown2020,Hartke2023};
computationally, advances in method development and 
softwares are opening a new era in the capabilities to study strongly correlated systems. This combination can be deployed for other models and for understanding many other problems in strongly correlated states of matter.

\vskip0.20in \noindent
\textit{Acknowledgements-}  
We are grateful to Jens Hertkorn for a critical reading of the manuscript. We thank Immanuel Bloch for helpful discussions. We thank the Flatiron Institute Scientific Computing Center for computational resources.
The experimental work at MIT was supported by the NSF Center for Ultracold Atoms and PHY-2012110, AFOSR (FA9550- 23-1-0402), DOE (DE-SC0024622), and the Vannevar Bush Faculty Fellowship (ONR N00014- 19-1-2631). C Turnbaugh was supported by the National Science Foundation Graduate Research Fellowship Program (NSF GRFP Grant No. 2141064).
Y-Y. He was supported by the National Natural Science Foundation of China (under Grants No.12247103 and No.12204377), and the Youth Innovation Team of Shaanxi Universities. The Flatiron Institute is a division of the Simons Foundation.

\bibliography{FFLO}

\clearpage
\centerline{\bf Supplemental Materials}
\vskip0.10in

\vskip0.10in \noindent
In these Supplemental Materials, we present additional details concerning (1) comparisons between DQMC and CP-AFQMC for various quantities at $\beta=3$; (2) the relation between the wave-vector of the FFLO order and magnetization $m$; (3) pairing and spin X/Y correlations on rectangle lattices with widths $L_y=4,6,8$ at different temperatures; (4) pair momentum distribution at different temperatures on a $20 \times 20$ lattice.

\begin{figure}[htbp]
 \includegraphics[width=1\columnwidth]{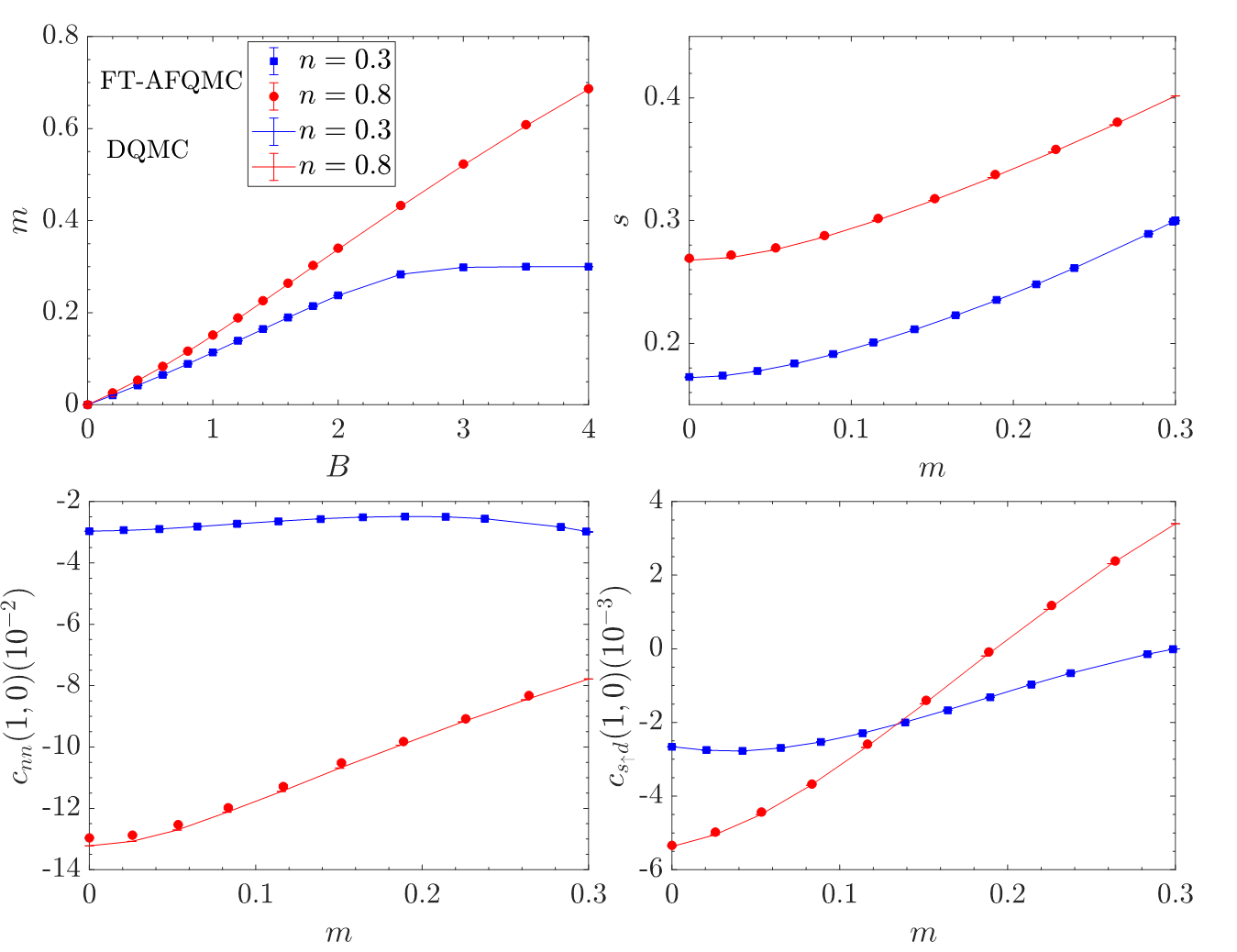}
\caption{\textbf{Comparisons between DQMC (curves) and CP-AFQMC (data points) at temperature $T=1/3$.} Magnetization of the system $m$, singlon density $s$, nearest neighbor density-density and singlon-up-doublon correlations as a function of magnetic field $B$ or magnetization $m$ at densities $n=0.3$ (blue) and $n=0.8$ (red). 
 The calculations are performed on a $16 \times 16$ square lattice at $U=-4$.}
 \label{fig:SM_comparison_DQMC_CPQMC}
 \end{figure}

{\it (1) Comparisons between DQMC and CP-AFQMC}
In Fig. ~\ref{fig:SM_comparison_DQMC_CPQMC}, we 
compare CP AFQMC  with DQMC at temperature $T=1/3$ on a $16 \times 16 $ lattice, where sufficiently small statistical error bars are still within reach for the latter. The top left panel shows magnetization $m$ vs.~magnetic 
field 
$B$ at fixed densities $n=0.3, 0.8$. Additional panels illustrate the singlon density $s=\frac{1}{N} \sum_{i}\langle s_{i\uparrow} + s_{i\downarrow}\rangle$, nearest neighbor density-density $c_{nn}$ and singlon-up-doublon $c_{s_{\uparrow}d}$ correlations as a function of $m$. DQMC and CP AFQMC exhibit good agreement across all these quantities.

{\it (2) Relation between the magnetization and FFLO order wave vector}

\begin{figure*}[htbp]
 \includegraphics[width=2\columnwidth]{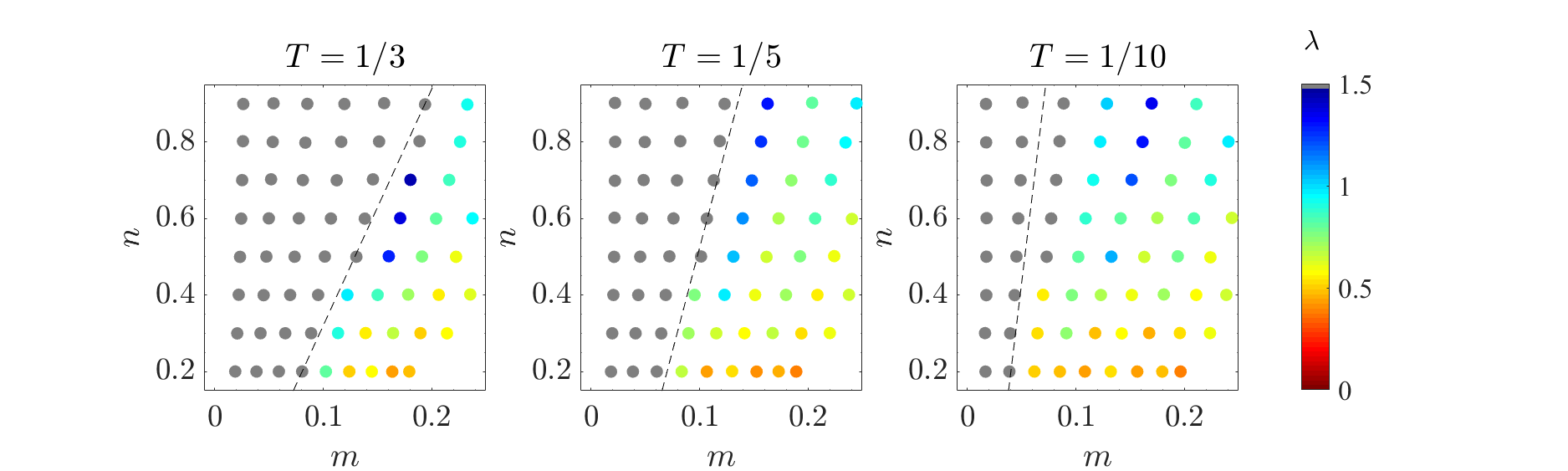}
\caption{\textbf{Relation between the magnetization $m$ and the FFLO wave vector $\left | \vec{Q} \right|$.} The color-bar values report the ratio $\lambda=m \pi / \left| \vec{Q} \right|$ on a magnetization-density plane at temperatures $T=1/3, 1/5, 1/10$, where $\vec{Q}$ is the peak position of the pair momentum distribution $n_{\vec{Q}}$ (refer to Fig. \ref{fig:signals_of_FFLO}). Grey color corresponds to $\lambda=\infty$, the non-FFLO regime where $\vec{Q}=(0,0)$. Calculations are performed on a $16 \times 16$ square lattice at $U=-4$.}
 \label{fig:SM_phase_diagram_ratio}
 \end{figure*}

In Fig.\ref{fig:SM_phase_diagram_ratio}, we report the ratio of magnetization $m$ to the length of the FFLO wave vector $\vec{Q}$ (the peak position of the pair momentum distribution $n_{\vec{Q}}$; see Fig.\ref{fig:signals_of_FFLO}), defined as $\lambda = m \pi / \left|\vec{Q}\right|$, on a density-magnetization plane at temperatures $T=1/3$, $T=1/5$, and $T=1/10$. The grey color corresponds to $\lambda = \infty$, indicating $\vec{Q} = (0,0)$ in the non-FFLO regime. Consistent with Hartree-Fock studies \cite{Chiesa2013}, $\lambda$ approaches 1 at relatively high densities and is slightly less than 1 at lower densities.

{\it (3) Pair \& Spin X/Y correlations at different temperatures on different lattices}

\begin{figure}[htbp]
 \includegraphics[width=1\columnwidth]{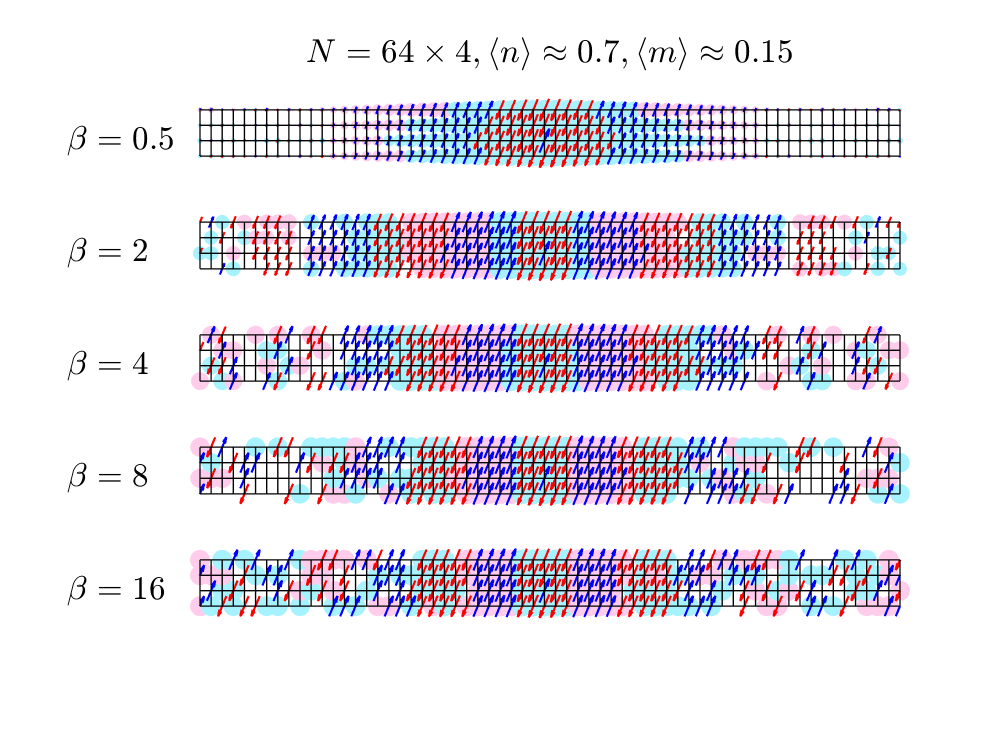}
\caption{Pairing and spin X/Y correlations at different temperatures on a rectangle $N=64 \times 4$ lattice with density $\langle n \rangle \approx 0.7$ and magnetization $\langle m \rangle \approx 0.15$. The cyan/pink circles represent positive/negative pairing correlations. The blue-up/red-down arrows represent positive/negative spin X/Y correlations. The length of the arrows stand for the relative magnitude of the correlations (log-scale). }
 \label{fig:pairing_spin_X_diff_lattice_size_64_4}
 \end{figure}

\begin{figure}[!b]
  \includegraphics[width=1\columnwidth]{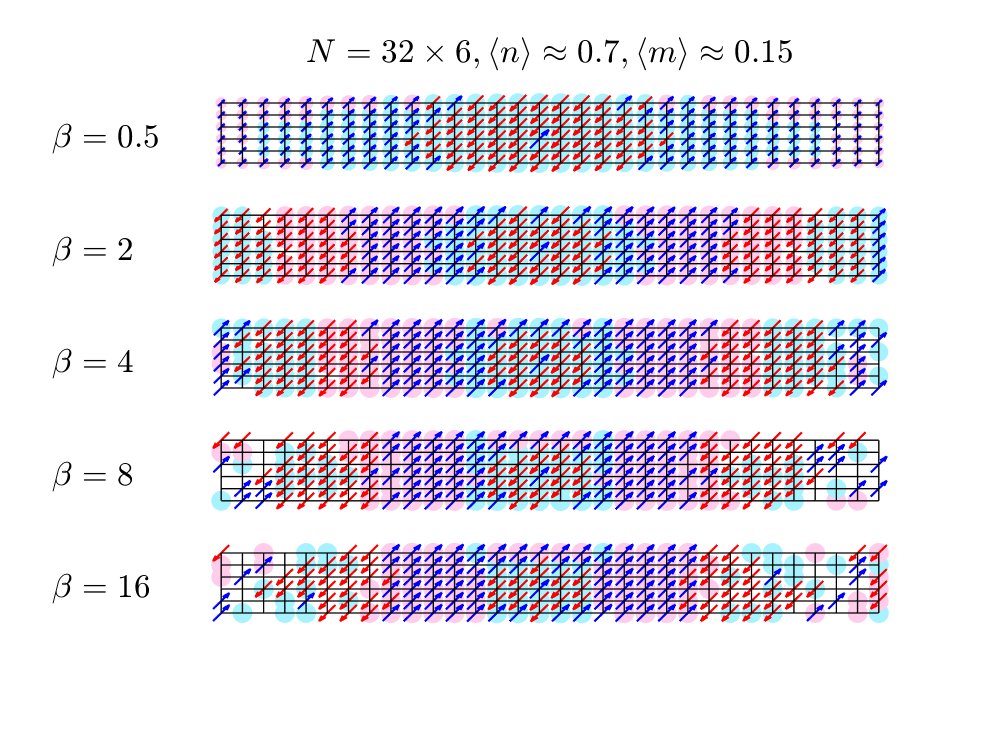}
\caption{Similar to Fig.~\ref{fig:pairing_spin_X_diff_lattice_size_64_4}, but for a $32 \times 6$ lattice.} \label{fig:pairing_spin_X_diff_lattice_size_32_6}
 \end{figure}

\begin{figure}[htbp]
   \includegraphics[width=1\columnwidth]{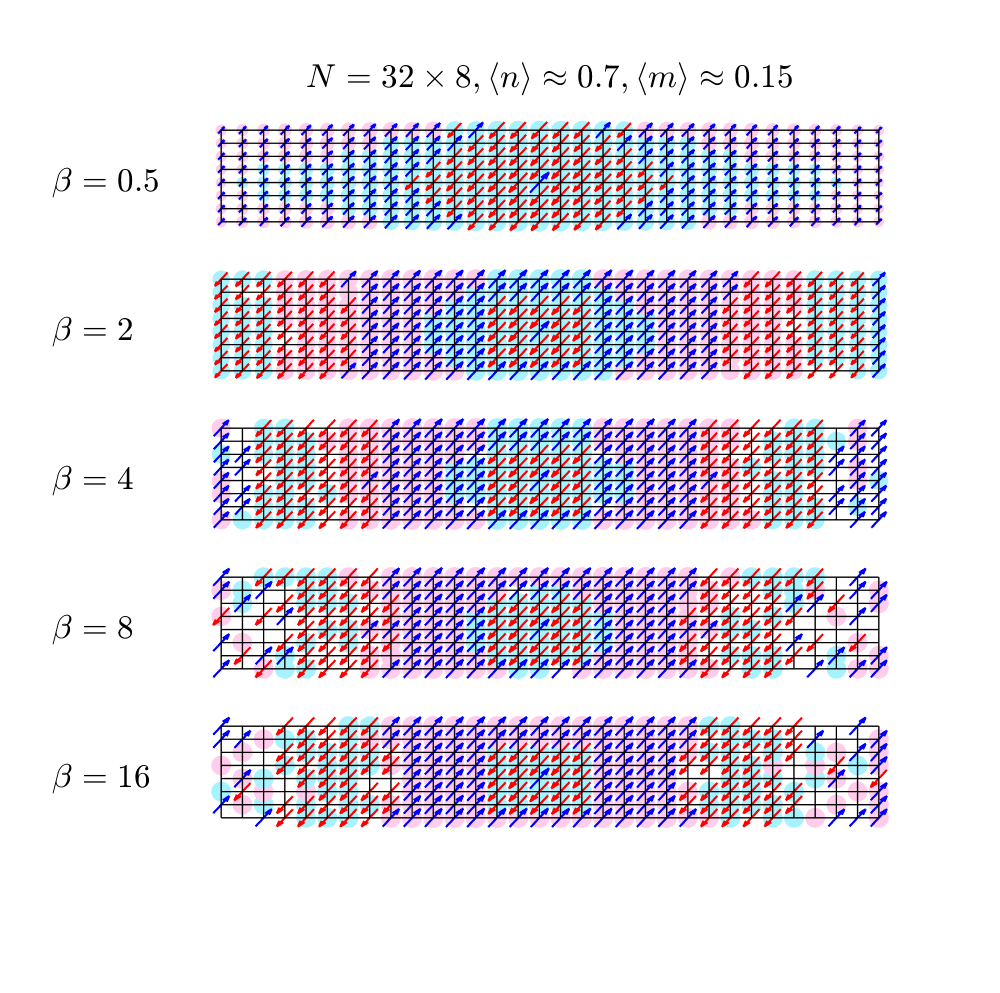}
\caption{Similar to Fig.~\ref{fig:pairing_spin_X_diff_lattice_size_64_4}, but for a $32 \times 8$ lattice.}
\label{fig:pairing_spin_X_diff_lattice_size_32_8}
 \end{figure}

We examine lattice size effects in Figs.~\ref{fig:pairing_spin_X_diff_lattice_size_64_4} through \ref{fig:pairing_spin_X_diff_lattice_size_32_8}, which show pairing and spin X/Y correlations on rectangular lattices of sizes $64 \times 4$, $32 \times 6$, and $32 \times 8$. Cyan/pink circles represent positive/negative pairing correlations, while blue-up/red-down arrows indicate positive/negative spin X/Y correlations. The same conclusions hold as those drawn from Fig.\ref{fig:pairing_spin_X}: at high temperatures, pairing correlations exhibit longer wavelengths than spin X/Y correlations. As the temperature decreases, the two become more correlated, as the
wavelengths of both correlations shorten and eventually become roughly equal at very low temperatures.

{\it (4) Pair momentum distribution at different temperatures on a $20 \times 20$ lattice.}

Here, we present 
the pair momentum distribution, $n_{\vec{Q}}$, as a function of $\vec{Q}$ at different temperatures on a $20 \times 20$ square lattice, similar to the right panel in Fig.~\ref{fig:signals_of_FFLO}. At high temperatures, the peak appears at $\vec{Q} = (0, 0)$, shifts to $(\pi/10, 0)$ as the temperature decreases to $T = 1/4$, and finally moves to $(\pi/5, 0)$ at $T = 1/16$.  
Thus, at low $T$ the peak position $(\pi/5, 0)$ lies between $(\pi/8, 0)$ and $(\pi/4, 0)$, the two allowed momenta on a $16 \times 16$ lattice, as shown in Fig.~\ref{fig:signals_of_FFLO}. Additionally, at $T = 1/16$, the value of $n_{\vec{Q}}$ at $(\pi/10, 0)$ (the first peak position at higher temperatures) is slightly lower than its value at higher temperatures, while the value at $(\pi/5, 0)$ (the peak position at extremely low temperatures) increases as the temperature decreases. As mentioned in the main text, this behavior explains the observation in Fig.~\ref{fig:signals_of_FFLO}, where $n_{\vec{Q}=(\pi/8, 0)}$ decreases slightly at low temperatures due to the incommensurate lattice size for $m = 0.15$.

  \begin{figure}[htbp]
   \includegraphics[width=1\columnwidth]{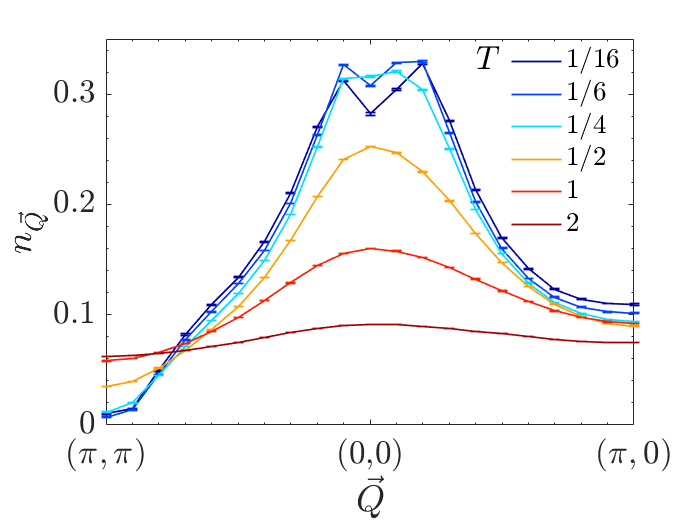}
\caption{Pair momentum distribution $n_{\vec{Q}}$ vs. $\vec{Q}$ on a $20 \times 20$ square lattice with $U=-4$, density $n=0.7$ and magnetization $m=0.15$.}
\label{fig:SM_N20_20_pair_momentum_distribution}
 \end{figure}

\end{document}